\documentclass[12pt]{article}

\def\rdots{\mathinner{\mkern1mu\raise1pt\vbox{\kern1pt\hbox{.}}\mkern2mu
\raise4pt\hbox{.}\mkern2mu\raise7pt\hbox{.}\mkern1mu}}
\newcommand{\Z}{{\rm Z\kern-.35em Z}}
\newcommand{\bP}{{\rm I\kern-.15em P}}
\newcommand{\Q}{\kern.3em\rule{.07em}{.65em}\kern-.3em{\rm Q}}
\newcommand{\R}{{\rm I\kern-.15em R}}
\newcommand{\h}{{\rm I\kern-.15em H}}
\newcommand{\C}{\kern.3em\rule{.07em}{.55em}\kern-.3em{\rm C}}
\newcommand{\T}{{\rm T\kern-.35em T}}

\begin{document}

\openup 1.5\jot

Submitted to Comm. Math. Phys. \ \ \ --- \ \ \ January 15, 1997

\bigskip
\centerline{REISSNER-NORDSTR\"{O}M-LIKE SOLUTIONS OF THE}
\centerline{$SU(2)$ EINSTEIN-YANG/MILLS EQUATIONS}

\bigskip
                    
\centerline{by J. A. Smoller and A.G. Wasserman}                    

\bigskip

In this paper we study a new type of solution of the  spherically
symmetric Einstein-Yang/Mills (EYM) equations with $SU(2)$ gauge group.
These solutions are well-behaved in the far-field, and have a
Reissner-Nordstr\"{o}m type essential singularity at the origin  $r =
0$.  These solutions display some novel features which are not present
in particle-like, or black-hole solutions.

     In order to describe these solutions and their properties, we
recall that 
for the spherically symmetric EYM equations, the Einstein metric is of
the 
form
$$ ds^2 = -AC^2 dt^2 + A^{-1} dr^2 + r^2 (d\theta^2 + \sin^2\theta
d\phi^2) 
\eqno(1.1)  $$
and the $SU(2)$ Yang-Mills curvature 2-form is
$$ F = w'\tau_1 dr \wedge d\theta + w'\tau_2  dr \wedge (\sin d\phi) -
(1-w^2)\tau_3 d\theta \wedge  (\sin \theta d\phi) \eqno(1.2). $$
Here  $A,C$, and  $w$  are functions of  $r$ ,  and  $\tau_1, \tau_2,
\tau_3$   form a basis for the Lie algebra  $su(2)$.  These equations
have been studied in many papers; 
see e.g. [3-19].

Smooth solutions of the EYM equations, defined for all  $r \ge 0$,  are
called (Bartnik-McKinnon, BM) particle-like solutions; such solutions
satisfy  
$1 > A(r) > 0$  for all $ r > 0$,  and  $A(0) = 1$.  The EYM equations
also admit 
black-hole solutions; i.e. solutions defined for all $r \ge  \rho > 0$,
where  $A(r) = 0$.
Here again, $1 > A(r) > 0$  for all  $r > \rho$.  The classical 
\vfill\eject
\noindent
Reissner-Nordstr\"{o}m (RN) solutions of the Einstein equations with
zero electric charge,  
$A(r) = 1 - {c \over r} + {1 \over r}, (c = const.)$,  $(AC^2)(r) =  (1
- {c \over r} + {1 \over r^2})$, are also solutions to the EYM
equations, with  $w(r) \equiv 0$.  We note that for this solution,
$A(r) > 1$  for  $r$  near  $0$.  We prove the existence of 
solutions which have this feature, $(A(r) > 1$  for $ r$  near  $0$),
of the 
classical RN solution, and we study their properties; we call these
Reissner- 
Nordstr\"{o}m-Like (RNL) solutions.$^1$

For these RNL solutions, we show that if  $A(r_1 ) = 1$  for some  $r_1
> 0$, 
then the solution is defined for all  $r$, $\  0 < r \le r_1$  , and
$\lim_{r\searrow 0} A(r) = \infty  = \lim_{r \searrow 0}(AC^2 )(r)$ .
However, the function  $B(r) \equiv r^2A(r)$ is analytic, on $0 \le r
\le r_1$  , as is the function  $w(r)$; moreover  $\lim_{r \searrow
0}w'(r) = 0$.

If we consider solutions that are defined in the far field; i.e., for
$r >> 1$, then it was shown in [10] that  $\lim_{r\searrow\infty}(A(r),
w^2 (r), w'(r)) = (1, 1, 0)$.
Thus the projection of the solution in the  $w - w'$ plane for a
particle like 
solution starts at the ``rest point" $(\pm 1,0)$, and goes to a ``rest
point" $(\pm 1,
0)$.  Black-hole solutions start at certain curves in the  $w - w'$
plane, 
([9]), and end at a rest point $(\pm 1,0)$.  In both of these cases,
there are an 
infinite number of solutions, distinguished by their nodal class
([8,9]).  
For RNL solutions, there is a parameter  $\sigma > 0$  defined by
$A(\sigma) = 1$.  We 
prove that for fixed  $\sigma$,  there are an infinite number of RNL
solutions 
distinguished by their integral nodal class, which must start at  $r =
0$  on 
the line $w' = 0$, and end at a rest point (? 1,0).  The RNL solutions
corresponding to the special case  $w(0) = 0$  are tangent to the line
$w = 0$; these 
give rise to half-integral nodal classes.

\bigskip
\noindent
\underline{\ \ \ \ \ \ \ \ \ \ \ \ \ \ \ }

$^1$ We base the name RNL on the behavior of such solutions near $r=0$.
For such solutions which are connecting orbits and have finite (ADM)
mass, the results in [2, p. 393] show that $A(r) = 1 - {c \over r} +
{0(1) \over r^2},$ as $r \rightarrow \infty$ and thus
behave differently at  $r = \infty$ from the RN solutions.  (The RN
solutions that 
we consider have zero electric charge and unit magnetic charge; c.f.
[18]. We 
thank P. Bizon for pointing this out to us.)
\vfill\eject

The proof of the existence of locally defined RNL solutions relies on a
local existence theorem at  $r = 0$, where we show that there is a
3-parameter 
family of analytic solutions starting at  $r = 0,  w'(0) = 0$.  The
proof of  
the existence of these local analytic solutions
 is non-trivial because the associated vector field is not even
continuous at  $r = 0$; (see the last part of $\S3$). Some of these
solution have been found numerically in [20].  It is 
interesting to note that when the first parameter  $w(0) = \pm 1$, and
the second 
parameter  $b_1  = 0$,  we recover the BM solutions.  If  $b_1  > 0$,
we get RNL 
solutions and if  $b_1  < 0$  we get Schwarzschild-like solutions.

We also prove that for fixed  $\sigma > {1 \over 2}$ , the (ADM) masses
of a sequence of 
our RNL solutions for  $\sigma$  fixed, and increasing nodal class,
tends to  ${1 \over \sigma}$.  Furthermore, the globally defined RNL
solutions which we obtain all have 
naked singularities at  $r = 0$; there may well be other RNL solutions
for 
which the singularity at  $r = 0$  is inside an event horizon.  We prove
that 
the singularity at  $r = 0$  for these RNL solutions is always
non-removable.

We show how the results which we have obtained, enable us to classify
all spherically symmetric EYM solutions, with  $SU(2)$ gauge group,
which are smooth and satisfy  $A > 0$  in the far field.

\bigskip
Please direct requests for hard copies of complete paper to
awass@math.lsa.umich.edu or smoller@math.lsa.umich.edu

\bigskip

\centerline{REFERENCES}                 

\bigskip
\begin{enumerate}
\item Adler, R., Bazin, M. Schiffer, M., Introduction to General
Relativity,
     2nd   ed., McGraw Hill, New York, (1975).
\item Arnowitt, R., Deser, S., Misner, C.W., The dynamics of general 
     relativity, in:  Gravitation, ed. by L. Witten, Wiley, N.Y.
227-265,  (1962).
\item   Bartnik, R., and McKinnon, J., Particle-like solutions of the 
     Einstein-Yang-Mills equations, Phys. Rev. Lett., 61, 141-144,
(1988).
\item   Bizon, P., Colored black holes, Phys. Rev. Lett. 64, 2844-2847,
(1990).
\item  Breitenlohner, P., Forg?cs, P., and Maison, D., Static
spherically 
     symmetric solutions of the Einstein-Yang-Mills equations, Comm.
Math. 
     Phys., 163, 141-172, (1994).
\item   Kunzle, H.P., and Masood-ul-Alam, A.K.M., Spherically symmetric
static  
     SU(2) Einstein-Yang-Mills fields, J. Math. Phys. 31, 928-935
(1990).
\item   Smoller, J., Wasserman, A., Yau, S.-T., McLeod, J., Smooth
static 
     solutions of the Einstein/Yang-Mills equations, Comm. Math. Phys.,
143, 
     115-147 (1991).
\item   Smoller, J., and Wasserman, A., Existence of infinitely-many
smooth 
     static, global solutions of the Einstein/Yang-Mills equations,
Comm. 
     Math. Phys., 151, 303-325, (1993).
\item   Smoller, J., Wasserman, A., and Yau, S.-T., Existence of black
hole 
     solutions for the Einstein-Yang/Mills equations, Comm. Math. Phys.,
154, 
     377-401, (1993).
\item  Smoller, J., and Wasserman, A., Regular solutions of the
Einstein-Yang/
     Mills equations, J. Math. Phys., 36, 4301-4323, (1995).
\item  Smoller, J. and Wasserman, A., Limiting masses of solutions of
Einstein- 
     Yang/Mills equations, Physica D, 93, 123-136, (1996).
\item  Smoller, J. and Wasserman, A., Uniqueness of extreme
Reissner-Nordstr?m 
     solution in  SU(2)  Einstein-Yang-Mills theory for spherically
symmetric 
     spacetime, Phys. Rev. D, 15 Nov. 1995, 52, 5812-5815, (1995).
\item  Straumann, N., and Zhou, Z., Instability of a colored black hole
solution, Phys. Lett. B. 243, 33-35, (1990).
\item  Smoller, J., and Wasserman, A., Uniqueness of zero surface
gravity SU(2) 
     Einstein-Yang/Mills black-holes, J. Math. Phys. 37, 1461-1484,
(1996).
\item  Smoller, J., and Wasserman, A., An investigation of the limiting
behavior of particle-like solutions to the Einstein-Yang/Mills equations
and a new black hole solution, Comm. Math. Phys., 161, 365-389, (1994).
\item Ershov, A.A., and Galtsov, D.V., Non abelian baldness of colored 
     black holes, Phys. Lett. A, 150, 747, (1989), 160-164.
\item Lavrelashvili, G., and Maison, D., Regular and black-hole
solutions 
     of Einstein/Yang-Mills dilaton theory, Phys. Lett. B, 295, (1992),
67.
\item Bizon, P. and Popp, O., No-hair theorem for spherical monopoles
and 
     dyons in SU(2) Einstein-Yang-Mills theory, Class.  Quantum Gravity,
9, 
     (1992), 193.
\item Volkov, M.S., and Gal'tsov, D.V., Sphalerons in
Einstein/Yang-Mills 
     theory.  Phys. Lett. B, 273, (1991), 273.
\item  Donets, E.E., Gal'tsov, D.V., and Zotov, M.Yu., Internal
structure of 
     Einstein-Yang-Mills black holes, gr-qc/9612067.
\end{enumerate}

\end{document}